\documentstyle[seceq,preprint,eclepsf]{jpsj}

\def\runtitle{Effects of Magnetic Field on Josephson Current in SNS System}
\def\runauthor{Yasushi {\sc Ishikawa} and Hidetoshi {\sc Fukuyama}}

\title
{Effects of Magnetic Field on Josephson Current \\in SNS System}
\author
{ 
Yasushi {\sc Ishikawa}\footnote{E-mail: ishikawa@watson.phys.s.u-tokyo.ac.jp}
and Hidetoshi {\sc Fukuyama}
}

\inst
{
Department of Physics, University of Tokyo, Tokyo 113
}

\recdate
{
September 19, 1998}

\abst
{The effect of a magnetic field on Josephson current has been studied
for a superconductor/normal-metal/superconductor (SNS) system, 
where N is a two-dimensional electron gas in a confining potential.
It is found that the dependence of Josephson currents on the magnetic field
are sensitive to the width of the normal metal.
If the normal metal is wide and contains many channels (subbands),
the current on a weak magnetic field shows a dependence similar to
a Fraunhofer-pattern and,
as the field gets strong, it shows another
type of oscillatory dependence on the field 
resulting from the Aharonov-Bohm interference between the edge states.
As the number of channels decreases ({\it i.e.} normal metal gets narrower),
however, the dependence in the region of the weak field deviates 
from a clear Fraunhofer pattern
and the amplitude of the oscillatory dependence 
in the region of the strong field is reduced.}

\kword
{SNS system,two-dimensional Normal metal, Fraunhofer pattern, 
Edge state, quantum wire}

\begin{document}
\sloppy
\maketitle
\newcommand{\vecvar}[1]{\mbox{\boldmath$#1$}}
\newcommand{\lsim}{ < \kern -11.8pt \lower 5pt \hbox{$\displaystyle \sim$}}
\newcommand{\gsim}{ > \kern -12pt   \lower 5pt   \hbox{$\displaystyle \sim$}}
\newread\epsffilein    
\newif\ifepsffileok    
\newif\ifepsfbbfound   
\newif\ifepsfverbose   
\newif\ifepsfdraft     
\newdimen\epsfxsize    
\newdimen\epsfysize    
\newdimen\epsftsize    
\newdimen\epsfrsize    
\newdimen\epsftmp      
\newdimen\pspoints     
\pspoints=1bp          
\epsfxsize=0pt         
\epsfysize=0pt         
\def\epsfbox#1{\global\def\epsfllx{72}\global\def\epsflly{72}%
   \global\def\epsfurx{540}\global\def\epsfury{720}%
   \def\lbracket{[}\def\testit{#1}\ifx\testit\lbracket
   \let\next=\epsfgetlitbb\else\let\next=\epsfnormal\fi\next{#1}}%
\def\epsfgetlitbb#1#2 #3 #4 #5]#6{\epsfgrab #2 #3 #4 #5 .\\%
   \epsfsetgraph{#6}}%
\def\epsfnormal#1{\epsfgetbb{#1}\epsfsetgraph{#1}}%
\def\epsfgetbb#1{%
%
%
\openin\epsffilein=#1
\ifeof\epsffilein\errmessage{I couldn't open #1, will ignore it}\else
%
%
   {\epsffileoktrue \chardef\other=12
    \def\do##1{\catcode`##1=\other}\dospecials \catcode`\ =10
    \loop
       \read\epsffilein to \epsffileline
       \ifeof\epsffilein\epsffileokfalse\else
%
%
          \expandafter\epsfaux\epsffileline:. \\%
       \fi
   \ifepsffileok\repeat
   \ifepsfbbfound\else
    \ifepsfverbose\message{No bounding box comment in #1; using defaults}\fi\fi
   }\closein\epsffilein\fi}%
%
%
\def\epsfclipon{\def\epsfclipstring{ clip}}%
\def\epsfclipoff{\def\epsfclipstring{\ifepsfdraft\space clip\fi}}%
\epsfclipoff
\def\epsfsetgraph#1{%
   \epsfrsize=\epsfury\pspoints
   \advance\epsfrsize by-\epsflly\pspoints
   \epsftsize=\epsfurx\pspoints
   \advance\epsftsize by-\epsfllx\pspoints
%
%
   \epsfxsize\epsfsize\epsftsize\epsfrsize
   \ifnum\epsfxsize=0 \ifnum\epsfysize=0
      \epsfxsize=\epsftsize \epsfysize=\epsfrsize
      \epsfrsize=0pt
%
%
     \else\epsftmp=\epsftsize \divide\epsftmp\epsfrsize
       \epsfxsize=\epsfysize \multiply\epsfxsize\epsftmp
       \multiply\epsftmp\epsfrsize \advance\epsftsize-\epsftmp
       \epsftmp=\epsfysize
       \loop \advance\epsftsize\epsftsize \divide\epsftmp 2
       \ifnum\epsftmp>0
          \ifnum\epsftsize<\epsfrsize\else
             \advance\epsftsize-\epsfrsize \advance\epsfxsize\epsftmp \fi
       \repeat
       \epsfrsize=0pt
     \fi
   \else \ifnum\epsfysize=0
     \epsftmp=\epsfrsize \divide\epsftmp\epsftsize
     \epsfysize=\epsfxsize \multiply\epsfysize\epsftmp   
     \multiply\epsftmp\epsftsize \advance\epsfrsize-\epsftmp
     \epsftmp=\epsfxsize
     \loop \advance\epsfrsize\epsfrsize \divide\epsftmp 2
     \ifnum\epsftmp>0
        \ifnum\epsfrsize<\epsftsize\else
           \advance\epsfrsize-\epsftsize \advance\epsfysize\epsftmp \fi
     \repeat
     \epsfrsize=0pt
    \else
     \epsfrsize=\epsfysize
    \fi
   \fi
%
%
   \ifepsfverbose\message{#1: width=\the\epsfxsize, height=\the\epsfysize}\fi
   \epsftmp=10\epsfxsize \divide\epsftmp\pspoints
   \vbox to\epsfysize{\vfil\hbox to\epsfxsize{%
      \ifnum\epsfrsize=0\relax
        \includegraphics{\ifepsfdraft}%
      \else
        \epsfrsize=10\epsfysize \divide\epsfrsize\pspoints
        \includegraphics{\ifepsfdraft}%
      \fi
      \hfil}}%
\global\epsfxsize=0pt\global\epsfysize=0pt}%
%
%
{\catcode`\%=12 \global\let\epsfpercent=
%
%
\long\def\epsfaux#1#2:#3\\{\ifx#1\epsfpercent
   \def\testit{#2}\ifx\testit\epsfbblit
      \epsfgrab #3 . . . \\%
      \epsffileokfalse
      \global\epsfbbfoundtrue
   \fi\else\ifx#1\par\else\epsffileokfalse\fi\fi}%
%
%
\def\epsfempty{}%
\def\epsfgrab #1 #2 #3 #4 #5\\{%
\global\def\epsfllx{#1}\ifx\epsfllx\epsfempty
      \epsfgrab #2 #3 #4 #5 .\\\else
   \global\def\epsflly{#2}%
   \global\def\epsfurx{#3}\global\def\epsfury{#4}\fi}%
%
%
\def\epsfsize#1#2{\epsfxsize}
%
%
\let\epsffile=\epsfbox

\section{Introduction}
 It is well known that in
a superconductor/insulator/superconductor (SIS) system, Josephson current 
as a function of a magnetic field shows a Fraunhofer pattern\cite{Tinkam}.
This remarkable phenomenon is caused by the interference effect 
of tunneling Cooper pairs. 

In the case of a SNS system, however, where N is a two-dimensional normal metal 
with a mesoscopic scale,
the dependences of the Josephson current on the magnetic field are expected to be not so
simple. 
For example, if the field is strong,
the normal metal becomes a quantum Hall system and the edge states are formed.
Ma and Zyuzin studied a SNS system under such a strong magnetic field \cite{MZ1}\cite{MZ2}
where N is two-dimensional normal metal confined by a linear potential  
and the edge states of the lowest Landau level are filled, 
and two superconducting leads
 are connected to the normal metal by point contact junctions.
They proposed that Josephson current flows through the edge states in the normal metal and
as a function of the field, the current shows Aharanov-Bohm type oscillations, 
similar to oscillations of the physical quantities in quantum dots. 

 In this paper, we study a SNS system where 
the normal metal is sandwiched between bulk superconductors.
Recently, such a structure has been actually constructed by Takayanagi 
{\it et al}\cite{Takayan}.
An external magnetic field penetrating the system can be arbitrary. 
The normal metal is assumed to be under the parabolic confining potential in the direction
perpendicular to the junction resulting in the discrete channels.
It turns out that the dependences of Josephson current are sensitive to 
the number of channels,
$N$, being defined as the number of channels in the absence of a magnetic field.

The organization of this paper is as follows.
In \S2, the model and a derivation of Josephson current are presented.
In \S3, we summarize the general features of the dependences of the current 
on a magnetic field and $N$.
\S4 and \S5 are devoted to a system with large $N$ under a weak field and
a strong field, respectively.
In \S6, we investigate how these results in a system with large $N$
is affected as $N$ decreases.
In \S7 we summarize these theoretical results and discuss its realizability.   

\section{Model and Derivation of the Josephson current}
We consider a system as shown in Fig.~\ref{System},
where two bulk superconductors are attached to a two-dimensional
normal metal under the confining potential. 
We assume that a magnetic field is perpendicular 
to the two-dimensional normal metal and penetrates only the normal metal region
uniformly. 

The Hamiltonian of the system is
\begin{equation}
\hat{\cal H } = \hat{\cal H }_{\mit S1}+ \hat{\cal H }_{\mit S2}
+\hat{\cal H}_{\mit N}+ \hat{\cal H}_{\mit Tunnel}\makebox[0.5em]{}, 
\end{equation}
where $ \hat{\cal H }_{\mit Si}(i=1,2)$ and $\hat{\cal H }_{\mit N }$
are the BCS Hamiltonian of two superconductors and 
the two-dimensional normal metal, respectively.

In the normal metal, the confining potential is assumed to be parabolic as
\cite{Yoshioka}\cite{Hajdu}
\begin{equation}
V(y)=\frac{1}{2}m\omega_0^2y^2 ,
\end{equation}
where $m$ is the electron mass, 
and the width of the normal metal 
in the $y$-direction, $W$, is classically defined as
\begin{equation}
\frac{1}{2}m\omega_0^2\left(\frac{W}{2}\right)^2\equiv\mu\makebox[0.5em]{},
\end{equation}
where $\mu$ is the chemical potential.
In the $x$-direction, electrons can move free along the length $L$.

The vector potential $\vecvar{A}$ is taken as 
$\vecvar{A}=-Hy\makebox[0.2em]{}\vecvar{e}_x$ in the normal metal, 
and $\vecvar{A}=\vec{0}$ in the superconductors. 
This choice of gauge enables us to regard the macroscopic phase of 
Cooper pairs in two superconductors as spatially uniform. 
In the calculation in this paper, for the sake of simplicity,
we neglect Zeeman energy whose effects
are briefly discussed in \S 7. 

For the model described above,
an eigenfunction in the normal metal is given by
\begin{equation}
\psi_{nk}(\vecvar{r})=\frac{e^{{\rm i}kx}}{\sqrt{L}}\cdot
\frac{1}{\sqrt{2^nn!\sqrt{\pi}l_0}}\cdot
\exp\left[-\frac{1}{2}\left(\frac{y-y_0(k)}{l_0}\right)^2\right]
\cdot H_n\left[\frac{y-y_0(k)}{l_0}\right],
\end{equation}
where $n=0,1,2,...$ and $k=0, \pm\pi/L, \pm 2\pi/L,... $ and
$H_n$ is the Hermite polynomial and
$\makebox[0.3em]{}y_0(k)=kl_0^4/l_c^2\makebox[0.3em]{}
,\makebox[0.5em]{}l_0=\sqrt{\hbar/m\omega}
\makebox[0.3em]{},\makebox[0.3em]{}l_c=\sqrt{\hbar/m\omega_c}
\makebox[0.3em]{},\makebox[0.3em]{}
\omega=\sqrt{\omega_0^2+\omega_c^2}\makebox[0.3em]{},\makebox[0.3em]{}
\omega_c=eH/mc$ being the cyclotron frequency.
We denote this eigenstate as $\{n,k\}$.

The eigenenergy $E_{n k}$ of the state $\{n,k\}$ is given by
\begin{equation}
E_{nk}=\hbar\omega\left(n+\frac{1}{2}\right)+\frac{\hbar^2k^2}{2M},
\end{equation}
where $M=m(\omega/\omega_0)^2$.

This spectrum is shown in Fig.~\ref{Spectrum}.
The Fermi wave number of each channel (subband, specified by $n$) is given by
\begin{eqnarray}
k_F(n)&=&\sqrt{\frac{2M}{\hbar^2}\left[\mu-\hbar\omega\left(n+\frac{1}{2}\right)
\right]}\nonumber\\
&=&K_F\cdot\sqrt{1-\frac{\hbar\omega}{\mu}\left(n+\frac{1}{2}\right)},
\end{eqnarray}
where $\hbar^2 K_F^2/2M=\mu$ 
( {\it i.e.} $K_F=W/2l_0^2$ ).
The chemical potential $\mu$ is determined to conserve the number of electrons 
in the normal metal under a varying magnetic field.
Hence, as the field increases, the number of occupied channels decreases.

The states around $\{n,\pm k_F(n)\}$ contribute to transport phenomena 
at low temperatures. 
The characteristics of the eigenfunction, $\psi_{n \pm k_F(n)}(\vecvar{r})$, 
are as follows\cite{Datta}.
The center of the wave function in the $y$-direction, $y_0(\pm k_F(n))$, is given by
\begin{equation}
y_0(\pm k_F(n))=\pm\frac{\omega_c}{\omega}\cdot\frac{W}{2}\cdot
\sqrt{1-\frac{\hbar\omega}{\mu}\left(n+\frac{1}{2}\right)}.
\end{equation}
The spatial extent of $\psi_{n \pm k_F(n)}(\vecvar{r})$ in the $y$-direction around 
$y_0(k_F(n))$ is roughly $l_0\sqrt{n+1/2}$. 
In the region of a weak magnetic field $(\omega_c/\omega_0\lsim\makebox[0.4em]{}1)$,
$y_0(k_F(n))$ is small compared with the classically defined edge $W/2$.
Therefore the currents carried by the states $\{n,k_F(n)\}$ and $\{n,-k_F(n)\}$ 
are almost canceled all over the normal metal.
As the field is increased $(\omega_c/\omega_0\gsim\makebox[0.4em]{}1)$, 
the factor $\omega_c/\omega$ in eq. (2.7) approaches to $1$.
So $y_0(\pm k_F(n))$ is written as
\begin{equation}
y_0(\pm k_F(n))\simeq\pm\frac{W}{2}\cdot
\sqrt{1-\frac{\hbar\omega}{\mu}\left(n+\frac{1}{2}\right)}
\makebox[1em]{} \left(\frac{\omega_c}{\omega_0}\gg 1\right),
\end{equation}
and spatial extent $l_0$ becomes small and converges to the Larmor radius $l_c$. 
Then the difference between $y_0(k_F(n))$ and $y_0(-k_F(n))$ becomes large,
and $y_0(\pm k_F(n))$ for a small $n$ almost coincides with the classically defined edge
$\pm W/2$. So the currents carried by $\{n,k_F(n)\}$ and $\{n,-k_F(n)\}$ are spatially 
separated and this leads to the formation of the edge states, that is to say,
the states carrying current along $+x$ shift to one side of 
the normal metal, while states carrying current in the other direction shift 
to the other side of the normal metal. 
These influences of a magnetic field on the eigenfunction 
are in accordance with a classical viewpoint, since the Lorentz force 
$(-e)\vecvar{v}\times\vecvar{B}$ is opposite for electrons moving in opposite directions. 

So far are discussed electronic states in the normal metal. 
The tunneling process between the superconductors and normal metal is described by
 $\hat{\cal H }_{\mit Tunnel}$ in eq. (2.1) as the following form,
\begin{equation}
\hat{\cal H }_{\mit Tunnel}= g\sum_{\sigma}\int d\vecvar{r}
\left\{\delta(x)\hat\psi_{S1\makebox[0.2em]{}\sigma}^{\dag}(\vecvar{r})
\hat\psi_{N\makebox[0.1em]{}\sigma}(\vecvar{r})+
\delta(x-L)\hat\psi_{N\makebox[0.1em]{}\sigma}^{\dag}(\vecvar{r})
\hat\psi_{S2\makebox[0.2em]{}\sigma}(\vecvar{r}) 
+ h.c\right\},
\end{equation}
where $\hat\psi_{Si\makebox[0.2em]{}\sigma}(\vecvar{r})$ and 
$\hat\psi_{N\makebox[0.1em]{}\sigma}(\vecvar{r})$ 
are the field operators of two superconductors and the normal metal, respectively.

Josephson current is given by the general formula as follows\cite{Al'tshuler} ,
\begin{equation}
J = \frac{2e}{\hbar}\frac{\partial\Omega }{\partial \theta}\makebox[0.5em]{},
\end{equation}
where $\Omega$ and $\theta$ are the thermodynamic potential and the phase 
difference between the two superconductors attached to the normal metal,
respectively. The dependence of $\Omega$ on $\theta$
 can be obtained by a perturbative expansion with respect to the pair amplitude of 
superconductors\cite{Kresin}. We evaluate Josephson current to the lowest order, 
which means considering the most simple process
of a Cooper pair propagation as shown in Fig.~\ref{Cooper}.
This approximation for the Josephson current
is valid under the condition $L\gsim\makebox[0.4em]{} \xi_c$, 
where $\xi_c$ is a coherence length in the normal metal,
representing a characteristic penetration length of a Cooper pair
into the normal metal, and given by $\xi_c=v_F/2\pi T$ in a clean limit. 
We evaluate the pair amplitude in the local approximation as
\begin{eqnarray}
\Delta_{i}(\vecvar{r}-\vecvar{r}',\tau-\tau')
&\equiv&
g^{2}<\hat\psi_{Si\makebox[0.2em]{}\uparrow}(\vecvar{r},\tau)
\hat\psi_{Si\makebox[0.2em]{}\downarrow}(\vecvar{r}',\tau')> \nonumber\\
&\simeq&
g^{2}<\hat\psi_{Si\makebox[0.2em]{}\uparrow}(\vecvar{r},\tau)
\hat\psi_{Si\makebox[0.2em]{}\downarrow}(\vecvar{r},\tau')>
\delta(\vecvar{r}-\vecvar{r}')\makebox[0.5em]{} ,
\end{eqnarray}
which means that two electrons of
a Cooper pair in the superconductors tunnel into the normal metal at the same point. 

Based on these approximations, we can express Josephson current by the
following form, 
\begin{eqnarray}
J &=& \frac{2e}{\hbar}(-{\rm i})T\sum_{\omega_{m}}|\Delta(\omega_{m},H)|^{2}
\int dy_l\int dy_r
\left\{ \makebox[0.5em]{}
\cal G\mit_{N\uparrow}(\vecvar{r}_l,\vecvar{r}_r;\omega_{m})
\cal G\mit_{N\downarrow}(\vecvar{r}_l,\vecvar{r}_r;-\omega_{m})e^{{\rm i}\theta}
\right.\nonumber \\ 
& & \makebox[14.5em]{}-\left.
\cal G\mit_{N\uparrow}(\vecvar{r}_r,\vecvar{r}_l;\omega_{m})
\cal G\mit_{N\downarrow}(\vecvar{r}_r,\vecvar{r}_l;-\omega_{m})e^{-{\rm i}\theta}
\makebox[0.5em]{} \right\}\makebox[0.5em]{},
\end{eqnarray}
where $\vecvar{r}_l=(0, y_l)$ and $ \vecvar{r}_r=(L, y_r)$ are the coordinates
at the interfaces between the superconductor and the normal metal,   
and $\cal G\mit_{N \sigma}$ is
a thermal Green's function in the normal metal.
In eq. (2.12), $|\Delta(\omega_m, H)|$ is defined as
\begin{equation}
|\Delta(\omega_m, H)|=g\cdot\frac{|\Delta(H)|}{\sqrt{\omega_m^2+|\Delta(H)|^2}},
\end{equation}
where $\Delta(H)$ is the order parameter in the superconductor
under a magnetic field $H$.
With the eigenfunction in eq. (2.4), $\cal G\mit_{N \sigma}$ can be written as
\begin{equation}
\cal G\mit_{N \sigma}(\vecvar{r},\vecvar{r}';\omega_{m})=
\sum_{nk}\frac{\psi_{nk}(\vecvar{r})\psi_{nk}^{*}(\vecvar{r}')}
{{\rm i}\makebox[0.2em]{} \omega_{m}-\xi_{nk}}
\makebox[3em]{}(\makebox[0.5em]{} \xi_{nk}= E_{nk}-\mu\makebox[0.5em]{} )
\makebox[0.5em]{}.
\end{equation}
By use of this Green's function, eq. (2.12) is transformed into 
\begin{eqnarray}
J&=&\frac{8e}{\hbar}T\sum_{\omega_{m}>0}|\Delta(\omega_{m},H)|^{2}
{\rm Re} \left[
\int dy_l\int dy_r
\cal G\mit_{N\uparrow}(\vecvar{r}_l,\vecvar{r}_r;\omega_{m})
\cal G\mit_{N\downarrow}(\vecvar{r}_l,\vecvar{r}_r;-\omega_{m})
\right]
\sin\theta \nonumber\\
&\equiv& J_c \sin\theta,
\end{eqnarray}
where we took the symmetries of the Green's function into account (see Appendix A).
This result coincides with a well-known form of Josephson current 
proportional to the sine of the phase difference,
while the higher order correction in this perturbation,
which is ignored here, gives rise to 
the higher harmonics of $\sin\theta$.

$J_c$ defined in eq. (2.15) can be calculated as follows ( see Appendix B ),
\begin{eqnarray} 
J_c&=&\frac{8e}{\hbar}T\sum_{\omega_m>0}|\Delta(\omega_m,H)|^2
{\rm Re}\sum_{n_{1}n_{2}}\int\frac{dk_{1}}{2\pi}
\int\frac{dk_{2}}{2\pi}
\frac{e^{-{\rm i} k_{1}L}}{{\rm i}\makebox[0.1em]{} \omega_{m}-\xi_{n_{1}k_{1}}}\cdot
\frac{e^{-{\rm i} k_{2}L}}{-{\rm i}\makebox[0.1em]{} \omega_{m}-\xi_{n_{2}k_{2}}}\cdot
\frac{1}{2^{n_{1}}n_{1}!\sqrt{\pi}}\cdot
\frac{1}{2^{n_{2}}n_{2}!\sqrt{\pi}}\nonumber \\
&\times&\left\{\int d\left(\frac{y}{l_{0}}\right)
H_{n_{1}}\left[\frac{y-y_{0}(k_{1})}{l_{0}}\right]\cdot
H_{n_{2}}\left[\frac{y-y_{0}(k_{2})}{l_{0}}\right]\cdot
\exp\left[-\frac{1}{2}\left(\frac{y-y_{0}(k_{1})}{l_{0}}\right)^{2}
-\frac{1}{2}\left(\frac{y-y_{0}(k_{2})}{l_{0}}\right)^{2}\right]
\right\}^{2}\nonumber\\
&=&\frac{8e}{\hbar}T\sum_{\omega_m>0}|\Delta(\omega_m,H)|^2
\sum_{n_{1}n_{2}}\left(\frac{2M}{\hbar^{2}}\right)^{2}\cdot
\frac{e^{-2b_1 L}}{\sqrt{a_1^2+b_1^2}}\nonumber\\ 
&\times&\int\frac{dq}{2\pi}\frac
{2(b_1 q+a_1 b_1+a_2 b_2)\cos\alpha-\left\{(q+a_1)^2-a_2^2-b_1^2+b_2^2
\right\}\sin\alpha}
{\left\{(q+a_1-a_2)^2+(b_1+b_2)^2\right\}
 \left\{(q+a_1+a_2)^2+(b_1-b_2)^2\right\}}
\cdot F_{n_{1}n_{2}}\left[\frac{ql_{0}^{3}}{2l_{c}^{2}}\right],
\end{eqnarray}
where $q=k_1-k_2$ (the relative wave number in the $x$-direction),
$\kappa_1=a_1+{\rm i}b_1,\kappa_2=-a_2+{\rm i}b_2
\makebox[0.2em]{}(a_i,b_i\geq 0)$ and
$\alpha=(q+2a_1)L-1/2\cdot
\tan^{-1}[\omega_m/(\mu-\hbar\omega(n_1+1/2))]$.\\
$\kappa_1$ and $\kappa_2$ are defined as
\begin{eqnarray}
\left\{ \begin{array}{cc}
\frac{\hbar^2 \kappa_1^2}{2M}\equiv\mu-\hbar\omega\left(n_1+\frac{1}{2}
\right)+{\rm i}\omega_m \\
\frac{\hbar^2 \kappa_2^2}{2M}\equiv\mu-\hbar\omega\left(n_2+\frac{1}{2}
\right)-{\rm i}\omega_m
\end{array}\right.
\makebox[3em]{}
(\makebox[0.5em]{}{\rm Im}[\kappa_1], {\rm Im}[\kappa_2]>0
\makebox[0.5em]{}) .
\end{eqnarray}
$F_{n_{1}n_{2}}$ is defined by
\begin{equation}
F_{n_{1}n_{2}}[z]
\equiv\frac{N!}{M!}(2z^{2})^{M-N}e^{-2z^{2}}
\left\{L_{N}^{(M-N)}[2z^{2}]\right\}^{2}
\makebox[2em]{}
\left\{ \begin{array}{cc}
M=\max\{n_{1},n_{2}\}\\
N=\min\{n_{1},n_{2}\}
\end{array}\right. ,
\end{equation}
where $L_{n}^{(\alpha)}$ is the Laguerre polynomial.

For channels $n_i$ satisfying 
$\makebox[0.5em]{}\mu-\hbar\omega(n_i+1/2)>0\makebox[0.5em]{} $,
$a_i$ and $b_i$ can be approximated as 
\begin{eqnarray}
\left\{ \begin{array}{ll}
a_i(\omega_m)\simeq k_F(n_i)\\
b_i(\omega_m)\simeq \frac{\omega_m}{\hbar v_F(n_i)}\makebox[1em]{}.
\end{array}\right. 
\end{eqnarray}
The damping term $\exp(-2b_1 L)$ in eq. (2.16) can be written as 
$\exp(-(2m+1)\cdot L/\xi_c(n_1))$
 where  
$\xi_c(n_1)=\hbar v_F(n_1)/2\pi T$.
Here we consider the case where $\xi_c(n_1)\lsim\makebox[0.4em]{} L$ 
for all occupied channels $n_1$,
therefore in the summation over the thermal frequency $\omega_m (>0)$,
the term corresponding to $m=0$ becomes dominant.   

\section{General Features of the Dependences on Magnetic Field and $N$ }
Fig.~\ref{Phase} indicates the general features of 
the dependence on the magnetic field for various $N$, the number of channels
at $H=0$,
which is related with the classically estimated width $W$ introduced in eq. (2.3)
as follows 
\begin{equation}
\hbar\omega\left(N+\frac{1}{2}\right)\simeq
\frac{1}{2}m\omega_0^2\left(\frac{W}{2}\right)^2 .
\end{equation}
Therefore the large $N$ corresponds to the wide normal metal.
As the field is increased, the number of channels
 decreases and finally becomes 1 as seen from Fig.~\ref{Phase}.   

The dependence is roughly classified into two regions; `` Fraunhofer pattern''
region and `` Edge current '' region. 
In `` Fraunhofer pattern '' region, Josephson current shows a dependence on the field
like a Fraunhofer pattern as in a SIS system and 
in `` Edge current '' region, the dependence shows
another type of oscillation due to the interference between
the edge states in the normal metal.

\section{Dependence in the Region of Weak Magnetic Field and Large $N$}
In this section, we assume that $N$ is large enough ($N \sim 100$) and 
consider the case of the weak field 
$(\omega_c/\omega_0\lsim\makebox[0.4em]{}1)$.
Since a weak magnetic field is expected not to affect so much an orbital motion
of electrons in the normal metal, 
Josephson current will show a dependence similar to a Fraunhofer-pattern 
in a SIS system.

 The numerical result is shown in Fig.~\ref{Weakmag}. 
In this calculation, parameters are chosen as 
$W=28\makebox[0.2em]{}\xi,\makebox[0.3em]{} L=3\makebox[0.2em]{}\xi\makebox[0.2em]{}, 
\makebox[0.3em]{}\xi_c=5\makebox[0.2em]{}\xi \makebox[0.3em]{}$ 
in the absence of the field
$(\makebox[0.2em]{}\xi_c=\hbar v_F/2\pi T,
\makebox[0.2em]{} v_F=\hbar K_F/M,
\makebox[0.2em]{}\xi=\sqrt{\hbar/m\omega_0}\makebox[0.2em]{}) $.
This choice of  $W$ leads to $N=98$, and 
$|\Delta(\omega_m,H)|$ the pair amplitude in superconductors 
is assumed to be unchanged by a magnetic field.

From this result, we see that, 
in a weak magnetic field $\omega_c/\omega_0\lsim\makebox[0.4em]{}1$,
the Josephson current shows a dependence similar to a Fraunhofer pattern
as $J_0\sin(\pi\Phi/\Phi_0)/(\pi\Phi/\Phi_0)$ 
with $\Phi$ being the magnetic flux in the normal metal and a flux quantum
$\Phi_0=hc/2e$, 
although the distance between successive maxima 
gradually increases as $\omega_c/\omega_0$ exceeds about 
$0.5 \makebox[0.2em]{}$.
By fitting this numerical result with the above formula of a Fraunhofer pattern,
we can estimate the effective width of the normal metal to be about
$26\makebox[0.2em]{}\xi$, while
the classically defined width $W$ at $H=0$ is $28\makebox[0.2em]{}\xi$. 

These numerical results are understood as follows based on eq. (2.16).
In such a weak field as $\omega_c/\omega_0\lsim\makebox[0.4em]{}1$,
the integrand in eq. (2.16) has a strong peak around $q=-(a_1+a_2)$
with the width $|b_1-b_2|$.
Other region of $q$ does not contribute to 
the integrated value so much, because of the oscillation. 
In order to understand the dependence on parameters qualitatively,  
we approximately replace this integration by the value at $q=-(a_1+a_2)$ 
multiplied by $|b_1-b_2|$. 
\begin{eqnarray} 
J_c&\simeq&\frac{8e}{\hbar}T|\Delta(\pi T,H)|^2\cdot {\rm Re}\left[
\int dy_l\int dy_r
{\cal G}_{N\uparrow}(\vecvar{r}_l,\vecvar{r}_r;\pi T)
{\cal G}_{N\downarrow}(\vecvar{r}_l,\vecvar{r}_r;-\pi T)
\right]\makebox[5em]{}\nonumber\\
&\sim& \frac{8e}{\hbar}T|\Delta(\pi T,H)|^2\cdot 
\frac{(-1)}{4\pi}\left(\frac{2M}{\hbar^2}\right)^2\nonumber\\
&&\times\sum_{n_1 n_2}\frac{e^{-2b_1 L}}{\sqrt{a_1^2+b_1^2}}\cdot
\frac{|b_1-b_2|}{b_1-b_2}\cdot
\frac{a_2\cos[(a_1-a_2)L]}{a_2^2+\left(\frac{b_1+b_2}{2}\right)^2}\cdot
F_{n_{1}n_{2}}\left[\frac{-(a_{1}+a_{2})l_{0}^{3}}{2l_{c}^{2}}\right]
\end{eqnarray}

 $F_{n_1 n_2}[z]$ has the maximum 
around $|z|=1/2\sqrt{2N}\cdot(M-N+1/2)\cdot\pi\gamma/2$ 
where $M=\max\{n_1,n_2\},N=\min\{n_1,n_2\}$ and 
$\gamma=0.6$ (for details, see Appendix C).
Therefore, in the summation over $n_1$ and $n_2$, 
contributions from terms satisfying the condition   
\begin{eqnarray}
\frac{l_0^3}{2l_c^2}(a_1+a_2)
&\simeq&\frac{1}{2\sqrt{2n_1}}\cdot
\left(n_2-n_1+\frac{1}{2}\right)\cdot\frac{\pi\gamma}{2}\nonumber\\
&\simeq&\frac{1}{2\sqrt{2n_1}}\cdot
\frac{\mu}{\hbar\omega}\cdot\frac{(a_1^2-a_2^2)}{K_F^2}
\cdot\frac{\pi\gamma}{2},
\end{eqnarray}
dominate.
For such pairs, the argument of the cosine in eq. (4.1) is given by 
\begin{equation}
(a_1-a_2)L=\pi\cdot
\frac{H\cdot L\cdot W_{{\rm eff}}(n_1)}{\Phi_0}
\end{equation}
where $\Phi_0=hc/2e$ and $W_{\rm eff}(n_1)$ is defined by
\begin{eqnarray}
W_{{\rm eff}}(n_1)&\equiv&\frac{2}{\pi\gamma}\cdot W
\sqrt{1+\left(\frac{\omega_c}{\omega_0}\right)^2}
\cdot\sqrt{\frac{\hbar\omega\cdot n_1}{\mu}}\nonumber\\
&\simeq&\makebox[0.3em]{}W\sqrt{1+\left(\frac{\omega_c}{\omega_0}\right)^2}
\cdot\sqrt{\frac{\hbar\omega\cdot n_1}{\mu}}.
\end{eqnarray}
$W_{{\rm eff}}(n_1)$ can be considered as an effective width of the channel $n_1$.
This effective width $W_{{\rm eff}}(n_1)$ is an increasing function of $n_1$ and
for the largest $n_1$
( $\hbar\omega(n_1+1/2)\simeq \mu$ ), $W_{{\rm eff}}(n_1)$  
approximately coincides with the classically estimated width $W$. 
From eq. (4.3), each maximum of the oscillation as a function of $H$ corresponds to
the change of the number of quantum flux $\Phi_0$ penetrating the region which 
electrons in the channel $n_1$ effectively occupy.
After taking the summation over $n_1$, we get a Fraunhofer pattern. 

In the absence of a magnetic field, because of the following relation
\begin{equation}
\lim_{H\rightarrow 0}F_{n_1 n_2}\left[\frac{ql_0^3}{2l_c^2}\right]=
\delta_{n_1 n_2},
\end{equation}  
the Cooper pairs consisting of electrons in the same channel as 
( $-k_{F}(n_1),k_{F}(n_1)$ )  
mainly contribute to Josephson current, and the Cooper pairs have a zero momentum 
as shown in Fig.~\ref{Change}(a).
The above analysis shows that,
as a magnetic field is applied in the region $\omega_c/\omega_0\lsim\makebox[0.4em]{}1$, 
the pair states like 
($-k_{F}(n_2),k_{F}(n_1)$) becomes effective where $\{n_1,n_2\}$ meets
the condition eq. (4.2), 
and the Cooper pair has a momentum, $p_F(n_1)-p_F(n_2)$ 
(see Fig.~\ref{Change}(b)). 
Therefore, it can be concluded that
as a magnetic field is applied the Cooper pairs consisting of electrons in different
channels begin to contribute and the pairs begin to have a nonzero momentum.
This change of the electronic states in the pairs contributing to the Josephson current
result in a Fraunhofer pattern.

 The fact that the Cooper pairs contributing to the current begin to have 
a finite momentum leads to the spatial separation of Cooper pairs
flowing in the opposite direction each other,
although there is a much spatial overlap under a weak magnetic field.
This tendency can be easily understood by considering that the center of 
the single-particle wave function in the $y$-direction is given by 
$y_0(k)=kl_0^4/l_c^2$, proportional to the wave number $k$ in the $x$-direction
as discussed in \S2.

 The increase of the difference in field between successive maxima at 
$\omega_c/\omega_0\gsim\makebox[0.4em]{}0.5$ in Fig.~\ref{Weakmag}
can be understood as follows.
The difference between $n_1$ and $n_2$ satisfying eq. (4.2)
becomes larger as the field becomes stronger. 
However, the number of channels in the normal metal is restricted to about 
$\mu/\hbar\omega$.  
Therefore as the field is increased, the number of Cooper pairs contributing to 
Josephson current decreases and
the contribution of the Cooper pairs with a small $n$ remains.
This leads to the increase of the distance between successive maxima
as is seen by eq. (4.3).
  
\section{Dependences in the Region of Strong Magnetic Field and Large $N$}
 We now consider the effect of a strong magnetic field 
$(\omega_c/\omega_0\gsim\makebox[0.4em]{}1)$.
The number of channels are assumed to be large as in \S4 ($N=98$).

The numerical result is shown in Fig.~\ref{Strongmag}.  
All parameters are the same as those in \S4 and 
the pair amplitude $|\Delta(\omega_m,H)|$ is assumed to be unchanged by the field. 
As the cyclotron frequency $\omega_c$ increases beyond $\omega_0$, 
Josephson current starts to exhibit another type of oscillation
which eventually dominates the Fraunhofer pattern in the strong field limit.
The distance between successive maxima is not so different from 
that of a Fraunhofer pattern.

Such behaviors can be understood as follows.
In eq. (2.16)
the peak at $q=-(a_1+a_2)$, which was important under the weak field,
does not contribute in the present case of $\omega_c/\omega_0\gsim 1$,
because $F_{n_1 n_2}[ql_0^3/2l_c^2]$ at $q=-(a_1+a_2)$ is so small. 
Instead, in the present case of $\omega_c/\omega_0\gsim\makebox[0.4em]{} 1$,
contributions from small $z$ is important 
and we approximate $F_{n_1 n_2}[z]$ as follows,
\begin{eqnarray}
F_{n_1 n_2}[z]\simeq         
\left\{ \begin{array}{rr}
\frac{Q[n_1]^2}{z^2+Q[n_1]^2}\makebox[2em]{}(n_1=n_2)\\
0\makebox[4em]{}(n_1\neq n_2)
\end{array}\right. \makebox[2em]{} 
[\makebox[0.5em]{} |z|\lsim\makebox[0.4em]{}Q[n_1]\makebox[0.5em]{}],
\end{eqnarray}				
where $Q[n_1]=1/2\sqrt{2n_1+1}\makebox[0.2em]{}$.
Then we can perform the $q$-integration in eq. (2.16) to obtain
\begin{eqnarray}
J_c&\simeq&
\frac{8e}{\hbar}T|\Delta(\pi T,H)|^2\left(\frac{M}{\hbar^2}\right)^2\sum_{n_1}
\left[\frac{e^{-2b_1 L}}{(a_1^2+b_1^2)}\cdot\frac{\left(\frac{c_1}{2a_1}\right)^2}
{1+\left(\frac{c_1}{2a_1}\right)^2}+
\frac{\cos\left[2a_1 L-\tan^{-1}\left(\frac{c_1}{2a_1}\right)\right]}
{a_1\sqrt{\left(a_1^2+b_1^2\right)
\cdot\left(1+\left(\frac{c_1}{2a_1}\right)^2\right)}}
\cdot\frac{e^{-(c_1+2b_1)L}}{1+\frac{2b_1}{c_1}}\right]\nonumber\\
&\simeq&\frac{8e}{\hbar}T|\Delta(\pi T,H)|^2\left(\frac{M}{\hbar^2}\right)^2
\sum_{n_1}\frac{\cos[2a_1 L]}{a_1\sqrt{a_1^2+b_1^2}}\cdot
\frac{e^{-(c_1+2b_1)L}}{1+\frac{2b_1}{c_1}} ,
\end{eqnarray}
where $c_1\equiv 2l_c^2/l_0^3\cdot Q[n_1]$ and 
we assumed the relation $c_1/2a_1\ll 1$ and $c_1 L\lsim\makebox[0.4em]{} 1$.

This evaluation implies that Josephson current is mainly carried by 
the Cooper pairs consisting of electrons
($k_F(n_1), k_F(n_1)$) (see Fig.~\ref{Change}(c)).
This tendency can be understood by considering that 
under such a strong magnetic field the center of the single-particle wave function
in the $y$-direction $y_0(k_F(n_1))$ 
gets close to the classically estimated edge $W/2$, 
and the spatial extent in the $y$-direction $l_0$ becomes small.
In this case, as seen from eq. (2.16), a main contribution comes from
states satisfying $n_1=n_2$ and $y_0(k_1)\simeq y_0(k_2)$, namely $k_1\simeq k_2$ 
since $y_0(k)\propto k$. 
Therefore Cooper pairs flow through the edge state in the normal metal
as schematically shown in Fig.~\ref{Edgecurrent}.

Each term in eq. (5.2) involves an oscillatory function of $H$, $\cos(2a_1L)$. 
This argument, $2a_1 L$, can be expressed as follows,
\begin{eqnarray}
2a_1L=\pi\cdot\frac
{H\cdot L\cdot 2y_0(a_1)\left(\frac{\omega}{\omega_c}\right)^2}
{\Phi_0},
\end{eqnarray}
where $2y_0(a_1)\simeq 2y_0(k_F(n_1))=y_0(k_F(n_1))-y_0(-k_F(n_1))$,
{\it i.e.} the distance between two edge states around $y=y_0(k_F(n_1))$ 
and $y=y_0(-k_F(n_1))$, and
the factor $(\omega/\omega_c)^2$ approaches to $1$ in the large $H$ limit. 
Therefore this argument corresponds to the number of
the flux quantum $\Phi_0$ surrounded by the edge states 
as shown in  Fig.~\ref{Edgecurrent}.
This oscillation results from the Aharonov-Bohm interference between the Cooper pairs,
and this situation coincides with the result by Ma and Zyuzin\cite{MZ1}\cite{MZ2};
 under a strong magnetic
field, the Josephson current is carried by Cooper pairs which consist of electrons 
having the same velocities and flow along the edge of the two-dimensional normal metal,
which leads to the Aharonov-Bohm type oscillations as in quantum dots. 

In this system with many channels, 
as a magnetic field is increased, Cooper pairs rearranges themselves
from $(-k_F(n_2),k_F(n_1))$ to $(k_F(n_1),k_F(n_1))$
in due order from Cooper pairs with a larger $n$.
This is why Josephson current starts to show
another type of oscillation around $\omega_c/\omega_0\simeq\makebox[0.4em]{}1$.
In the strong magnetic field limit, almost all the Cooper pairs finally flow 
through the edge states. This leads to a complicated oscillation because
the amplitude and the periodicity of the oscillatory behavior in each channel 
are not correlated as is seen in eq. (5.2). 
However the distance between successive maxima of Josephson current are not so different
between weak and strong field limits, 
because the argument in eq. (5.3) is approximately the same form as one in eq. (4.3), 
although under a high magnetic field the chemical potential $\mu$ decreases
(\it i.e. \rm the width $W$ decreases).

\section{Quantum Limit ( Small $N$ )}
In \S4 and \S5, we assumed that there are many channels $(N\sim 100)$
in the normal metal, {\it i.e.} the width $W$ is large enough.
In this section, 
we consider the case where there are not so many channels in the normal metal.

The numerical results for several choices of $W$ are shown in Fig.~\ref{SmallN}.
In order to see the dependence on $N$ clearly, the width $W$
in Fig.~\ref{SmallN}(a) is chosen as $20.1\xi$ corresponding to $N=51$ 
which can be considered in the middle of a large $N$ and a small $N$. 
Here, $|\Delta(\omega_m,H)|$ is assumed to be unchanged by the field as before.
As $N$ decreases, we see an appreciable deviation from a Fraunhofer pattern
in the weak field region $(\omega_c\lsim\makebox[0.5em]{}\omega_0)$
and a less oscillatory behavior due to the edge states in the strong field region
$(\omega_c\gsim\makebox[0.5em]{}\omega_0)$.

In order to understand these properties, we consider a one-channel system $(N=1)$ 
as a limiting case, where the Josephson current is given 
\begin{equation}
J_c\simeq\frac{8e}{\hbar}T|\Delta(\pi T,H)|^2\left(\frac{M}{\hbar^2}\right)^2
\left[\frac{e^{-2b_1 L}}{(a_1^2+b_1^2)}\cdot\frac{\left(\frac{c_1}{2a_1}\right)^2}
{1+\left(\frac{c_1}{2a_1}\right)^2}+
\frac{\cos\left[2a_1 L-\tan^{-1}\left(\frac{c_1}{2a_1}\right)\right]}
{a_1\sqrt{\left(a_1^2+b_1^2\right)
\cdot\left(1+\left(\frac{c_1}{2a_1}\right)^2\right)}}
\cdot\frac{e^{-(c_1+2b_1)L}}{1+\frac{2b_1}{c_1}}\right],
\end{equation}  
where $c_1\equiv 2l_c^2/l_0^3\cdot Q[n_1]$.
Here we approximated
$F_{n_1 n_1}[z]$ in eq. (2.16) by the Lorentzian as in eq. (5.1).
 
Since the factor $c_1/2a_1$ is larger than $1$ in the region  
$\omega_c/\omega_0\lsim\makebox[0.4em]{}1$ 
$(c_1/2a_1=\infty$ when $ \omega_c\rightarrow 0)$, the first term dominates, 
which means that Josephson current does not show any oscillatory dependence in the 
region of the weak field. As shown in Fig.~\ref{Change}(b),
the Fraunhofer pattern is caused by Cooper pairs consisting of 
electrons belonging to the different channels.
Therefore as $N$ decreases, the interference between Cooper pairs is weakened.

Under the strong field $(\omega_c/\omega_0\gsim\makebox[0.4em]{}1)$, 
the factor $c_1/2a_1$ becomes smaller than $1$, and the 1st term becomes small.
However the factor $c_1 L$ in the damping factor $\exp(-(c_1+2b_1)L)$ in
the second term
is large for a small $n_1$, so the second term is much smaller than 
the first term and the oscillatory behavior does not appear.

Especially in Fig.~\ref{SmallN}(b) and (c), 
the decrease of the number of channels as the field increases affects Josephson
current. This is because the contribution of each channel to Josephson current becomes 
large in such a narrow system.

It can be concluded that as the number of channels decreases 
the field dependence in the weak field region 
$(\omega_c/\omega_0\lsim\makebox[0.4em]{}1)$
deviates from a Fraunhofer pattern 
and the amplitude of the oscillatory behavior due to the edge states 
under the strong field 
$(\omega_c/\omega_0\gsim\makebox[0.4em]{}1)$ is reduced. 

\section{Summary and Discussion} 
We have studied the effect of an external magnetic field on Josephson current in a SNS 
system where two bulk superconductors are attached to 
a confined two-dimensional normal metal 
and the temperature is assumed to satisfy $\xi_c\lsim\makebox[0.4em]{}L$
($\xi_c$ is the coherence length of the pair in the normal metal).   
We first studied the case that the normal metal has many channels $(N\sim 100)$.
In the absence of field,
Cooper pairs which consist of two electrons 
in the same channel and have an essentially zero momentum contribute 
to the current. 
For the weak field $(\omega_c/\omega_0\lsim\makebox[0.4em]{}1)$,
however, the contribution of pairs 
which consist of electrons in  the different channels 
and have the finite momentum $p_F(n_1)-p_F(n_2)$ becomes larger. 
This leads to a Fraunhofer pattern. 
As the field is increased further ($\omega_c/\omega_0\gsim\makebox[0.4em]{}1$),
Cooper pairs with a larger $n$ 
rearrange themselves into the states that
two electrons are in the same channel and have the same momentum.
This means that Josephson current flows through the edge states, and shows
an oscillatory dependence due to the Aharonov-Bohm interference between Cooper pairs.
This is physically the same as the result by Ma and Zyuzin\cite{MZ1}\cite{MZ2},
who studied the case where two superconducting leads are attached via point contacts
to a confined two-dimensional normal metal and the edge states of 
the lowest Landau level are filled.

We next examined the dependence on $N$ the number of channels in the normal metal at
$H=0$. The decrease of $N$ weakens
an interference between Cooper pairs even in the region of
 the weak field and leads to 
a deviation from a Fraunhofer pattern. 
In a limiting case, Josephson current in an one-channel system $(N=1)$
does not show a Fraunhofer pattern dependence at all.
Under the strong field, the amplitude of 
the oscillatory behavior due to the edge states is reduced as $N$ decreases.

In this paper, we neglected Zeeman energy which splits
a Fermi wave number of the each channels as $k_F+\Delta k, k_F-\Delta k$.
Under the strong field, the single-particle wave function $\psi_{n k}(\vecvar{r})$
is localized around the center $y_0(k)$ in the $y$-direction with the spatial extent
$l_c\sqrt{n+1/2}$. 
In the absence of the Zeeman energy, a much 
contribution comes from the pair states where two electrons belong to the same channel and
the centers of two single particle 
wave functions in the $y$-direction are in an almost same position. 
The Zeeman energy separates this overlapping 
two single-particle wave functions in the $y$-direction 
as $y_0(k_F+\Delta)$ and $y_0(k_F-\Delta)$.  
Therefore Zeeman energy will reduce Josephson current.  

In this paper, we assumed that the confining potential caused by the gate voltage 
is parabolic.
The self-consistent calculation for a GaAs/AlGaAs heterostructure
by Laux {\it et al}\cite{Laux} suggests that 
the parabolic potential 
is appropriate for a system with one or two channels ($N=1$ or $N=2$),
but as charge accumulates in the well
({\it i.e.} $N$ is increased in the region ), 
the potential gets close to a square well type.
If this is the case,
the behavior of Josephson current
in the system with a few channels as discussed in \S 6,
can be considered rather realistic.
In the system with a large $N(\sim 100)$ as discussed in \S4 and \S5, 
however, the edge states may be located
nearer to the classically estimated edge $\pm W/2$ regardless of $n$.
Therefore the periodicities of the oscillations due to edge states in eq. (5.3)
can be almost the same for all channels, and the oscillatory dependence due to edge states
is expected to become simpler than one shown in Fig.~\ref{Strongmag}. 
On the other hand,
Fraunhofer pattern under the weak field is considered not to be affected
so much by this alternation of a confining potential.
From these discussions, it can be expected that
the field dependence in a strong field region will reflect
the actual shape of the confining potential. 

\section*{Acknowledgment} 
Y.I would like to express his gratitude to Hiroshi Kohno, Ken Yokoyama, 
Masakazu Murakami and Flordivino Basco for instructive and useful discussions 
and careful reading of the manuscript,
and he is grateful to other colleagues in the Fukuyama laboratory.

\appendix 
\section{Symmetries of the Green's Function}
In the derivation eq. (2.15), we have used the following symmetries of 
the Green's functions; 
\begin{eqnarray}
&&\cal G\mit_{N \sigma}(\vecvar{r}_l,\vecvar{r}_r;-\omega_{m})
=\cal G\mit_{N \sigma}(\vecvar{r}_r,\vecvar{r}_l;\omega_{m})^{*}\\ 
&&\int dy_l\int dy_r \makebox[0.5em]{}
{\cal G}_{N\uparrow}(\vecvar{r}_l,\vecvar{r}_r;\omega_{m})
{\cal G}_{N\downarrow}(\vecvar{r}_l,\vecvar{r}_r;-\omega_{m})
\nonumber\\
&=&\int dy_l\int dy_r \makebox[0.5em]{}
{\cal G}_{N\uparrow}(\vecvar{r}_r,\vecvar{r}_l;\omega_{m})
{\cal G}_{N\downarrow}(\vecvar{r}_r,\vecvar{r}_l;-\omega_{m})\makebox[2em]{}
\end{eqnarray}

\section{Calculation of the Green's Function}
By use of the formula\cite{Table},
\begin{eqnarray}
\int_{-\infty}^{\infty}dx\makebox[0.5em]{} 
e^{-x^2}H_m[x+y]H_n[x+z]=2^n\pi^{\frac{1}{2}}m!z^{n-m}
L_m^{(n-m)}[-2yz]\makebox[3em]{}  (n\geq m),
\end{eqnarray}
we can transform $J_c$ as the following, 
\begin{eqnarray} 
J_c&=&\frac{8e}{\hbar}T\sum_{\omega_m>0}|\Delta(\omega_m,H)|^2{\rm Re}
\int dy_l\int dy_r
{\cal G}_{N\uparrow}(\vecvar{r}_l,\vecvar{r}_r;\omega_{m})
{\cal G}_{N\downarrow}(\vecvar{r}_l,\vecvar{r}_r;-\omega_{m})\nonumber\\
&=&\frac{8e}{\hbar}T\sum_{\omega_m>0}|\Delta(\omega_m,H)|^2
{\rm Re}\sum_{n_{1}n_{2}}2^{3}\left(\frac{2M}{\hbar^{2}}\right)^{2}
\int\frac{dq}{2\pi}\makebox[0.5em]{}
F_{n_{1}n_{2}}\left[\frac{ql_{0}^{3}}{2l_{c}^{2}}\right]
\nonumber\\ 
&\times&\int\frac{dK}{2\pi}\frac{e^{-{\rm i}KL}}
{(K+q-2\kappa_{1})(K+q+2\kappa_{1})(K-q-2\kappa_{2})(K-q+2\kappa_{2})},
\end{eqnarray}
where $K=k_1+k_2$(the total momentum), $q=k_1-k_2$(the relative momentum),
and $k_1$ and $k_2$ are treated as continuous here by the following reason.
At the interfaces between the normal metal and the superconductor, 
there will be a repulsive potential, {\it i.e.} potential barrier, in the $x$-direction,
through which electrons in the normal metal region are coupled to the propagating
states in the bulk superconductors. 
After performing the K-integration, we obtain eq. (2.16).
 
\section{Estimation of Maximum Point of $F_{n_1 n_2}[z]$}

The asymptotic behavior of the Laguerre polynomial\cite{Table} is given by
\begin{equation}
L_n^{(\alpha)}(x)=\frac{1}{\sqrt{\pi}}e^{\frac{x}{2}}
x^{-\frac{\alpha}{2}-\frac{1}{4}}n^{\frac{\alpha}{2}-\frac{1}{4}}
\cos\left[2\sqrt{nx}-\left(\alpha+\frac{1}{2}\right)\frac{\pi}{2}\right]
+O(n^{\frac{1}{2}\alpha-\frac{3}{4}}).
\end{equation}
which is adequate for a large $n$ and roughly\\
$1/2\sqrt{n}\cdot(\alpha+1/2)\cdot\pi/2\lsim\makebox[0.4em]{} \sqrt{x}
\lsim \makebox[0.4em]{} 1/2\sqrt{n}\cdot(\alpha+1/2)\cdot\pi/2+\sqrt{n}\pi/2$.\\
By use of this, $F_{n_1 n_2}[z]$ is approximated by
\begin{equation}
F_{n_1 n_2}[z]\simeq
\frac{1}{\pi\sqrt{N\cdot 2z^2}}
\cos^2\left[2\sqrt{N\cdot 2z^2}-\left(M-N+\frac{1}{2}\right)\cdot
\frac{\pi}{2}\right],
\end{equation}
where $M=\max\{n_1,n_2\},N=\min\{n_1,n_2\}$.

From this asymptotic form, 
$F_{n_1 n_2}[z]$ can be expected to have a maximum around 
$|z|=1/2\sqrt{2N}\cdot(M-N+1/2)\cdot\pi/2$.
However, this asymptotic form is valid in the larger region than this maximum point.
Therefore the real maximum point deviates from the expected one and can be written as
\begin{equation}
|z|=\frac{1}{2\sqrt{2N}}\cdot\left(M-N+\frac{1}{2}\right)\cdot\frac{\pi\gamma}{2},
\end{equation} 
where $\gamma$ is weakly dependent on $n_1, n_2$ and ranges roughly 
from $0.5$ to $0.7$. 
For simplicity, we have used the average value of $0.6$ for $\gamma$.

\begin{figure}
\caption{Schematic illustration of SNS system considered in this paper.
The origin of the coordinate system $(x,y)$ is chosen at the center of 
the left SN interface.}
\label{System}
\end{figure}

\begin{figure}
\caption{Dispersion relation, $E_{n k}$ vs $k$ for electronic subbands
arising from electronic confinement in zero magnetic field.}
\label{Spectrum}
\end{figure}

\begin{figure}
\caption{The lowest order process of Cooper pair propagation considered
in this paper. Electronic state in the normal metal is specified by $\{n,k\}$.}
\label{Cooper}
\end{figure}

\begin{figure}
\caption{Schematic picture showing the dependence of characteristic
magnetic field effects on $N$ the number of channels 
in the absence of the field.}
\label{Phase}
\end{figure}

\begin{figure}
\caption{The dependence of Josephson current on a weak magnetic field
($\omega_c/\omega_0\lsim\makebox[0.4em]{}1$) in a system with large $N$
(N=98). The parameters are chosen as 
$W=28\makebox[0.2em]{}\xi,L=3\makebox[0.2em]{}\xi,
\xi_c=5\makebox[0.2em]{}\xi$
($\xi=\sqrt{\hbar/m\omega_0}$). $J_c$ is normalized by 
$8emT/\hbar^3\mu\cdot|\Delta(\pi T,H)|^2\equiv J_0$.}
\label{Weakmag}
\end{figure}

\begin{figure}




\caption{Cooper pairs giving a dominant contribution to
Josephson current. (a)zero magnetic field, (b)weak magnetic field,
(c)strong magnetic field.}
\label{Change}
\end{figure}

\begin{figure}
\caption{Josephson current as a function of magnetic field
in the case of strong magnetic field
($\omega_c/\omega_0\gsim\makebox[0.4em]{}1$).
All parameters are the same as those in Fig.~\ref{Weakmag}, {\it i.e.}
$W=28\makebox[0.2em]{}\xi, L=3\makebox[0.2em]{}\xi,
\xi_c=5\makebox[0.2em]{}\xi$
($\xi=\sqrt{\hbar/m\omega_0}$)
}
\label{Strongmag}
\end{figure}

\begin{figure}
\caption{Schematic illustration of the motion of Cooper pairs flowing 
through the edge state under a strong magnetic field.}
\label{Edgecurrent}
\end{figure}

\begin{figure}
\caption{Josephson current as a function of magnetic field
for various choice of system width, $W$.
(a) $W=20.1\makebox[0.2em]{}\xi$ $(N=51)$, 
(b) $W=9.2\makebox[0.2em]{}\xi$ $(N=11)$,
(c) $W=5.3\makebox[0.2em]{}\xi$ $(N=4)$, 
(d) $W=2.2\makebox[0.2em]{}\xi$ $(N=1)$, 
where 
$L=3\makebox[0.2em]{}\xi, \xi_c=5\makebox[0.2em]{}\xi,
\xi=\sqrt{\hbar/m\omega_0}$. }
\label{SmallN}
\end{figure}

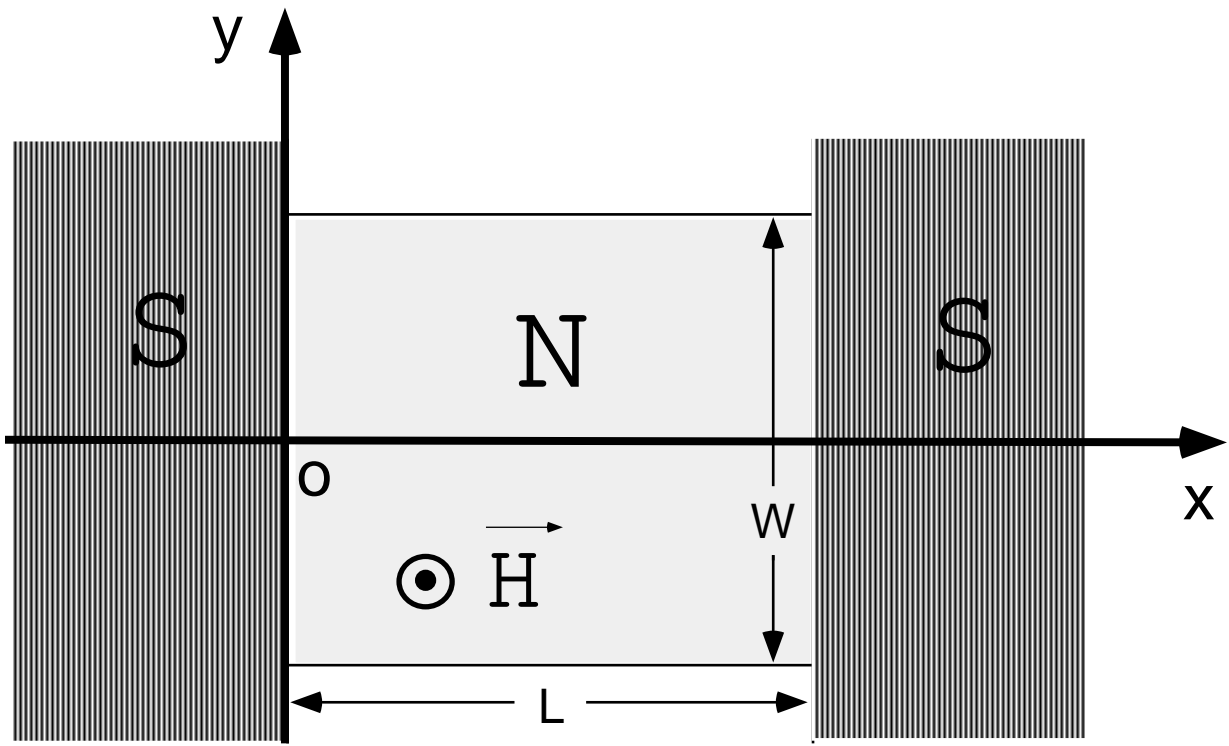
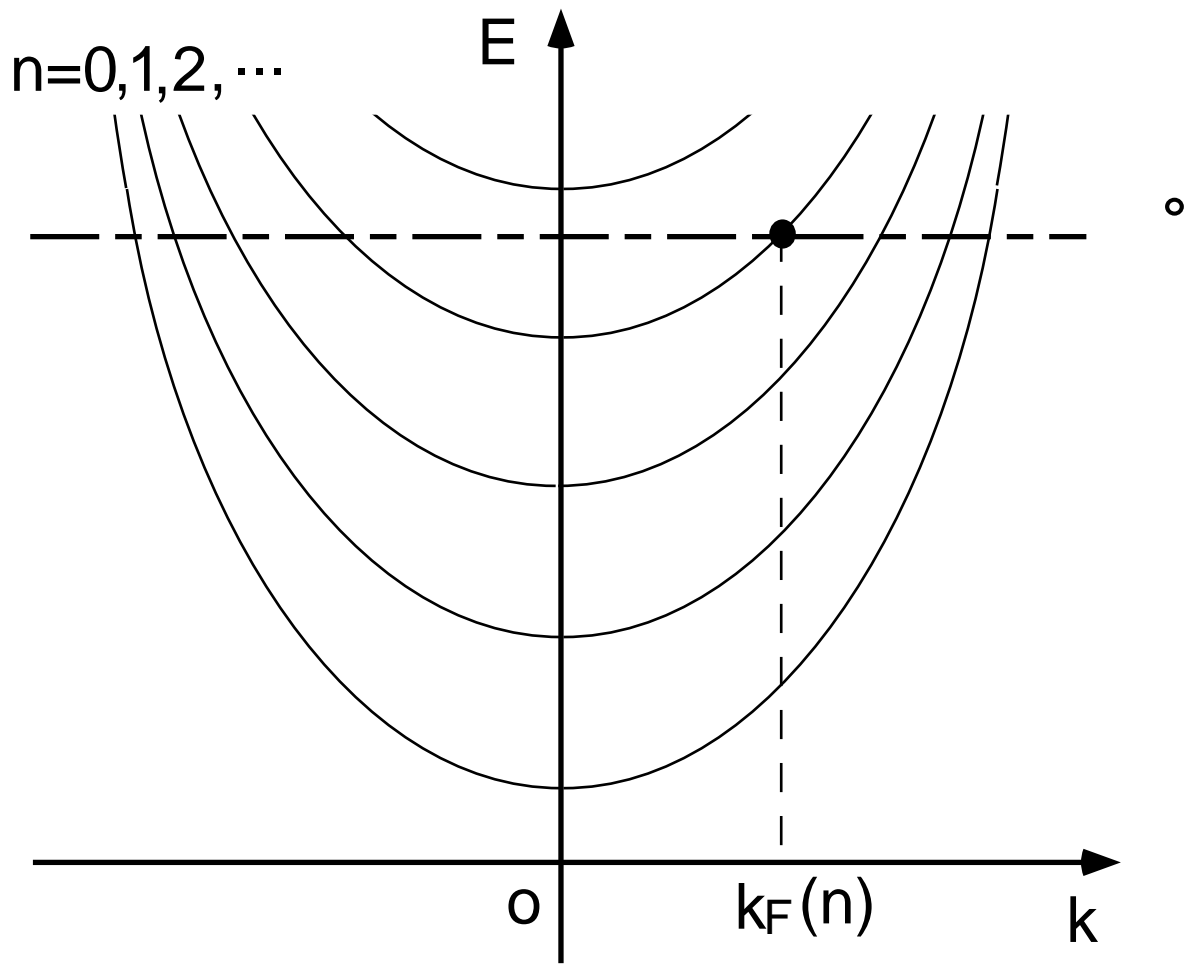
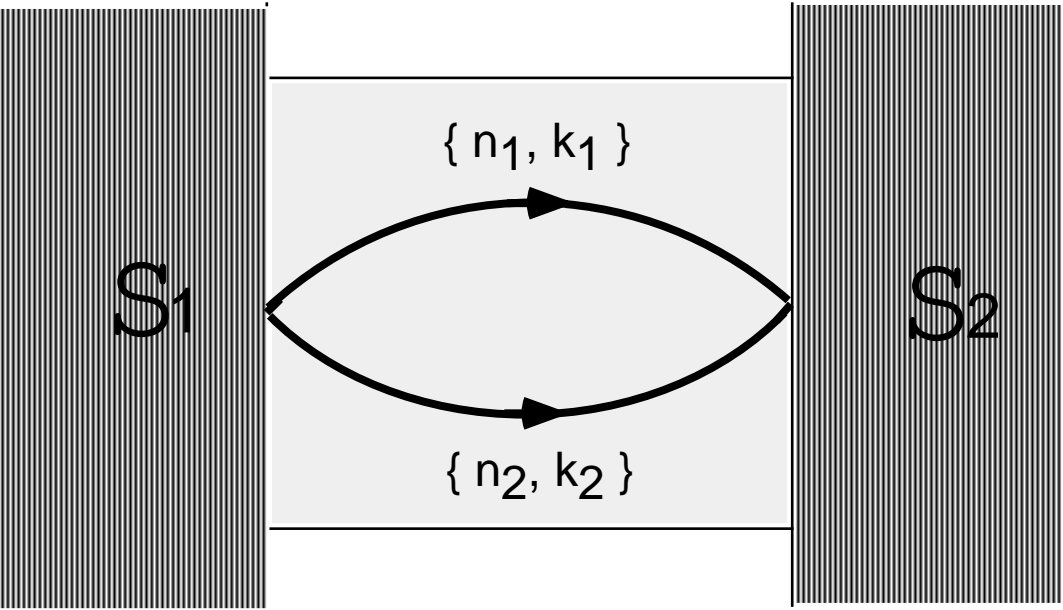
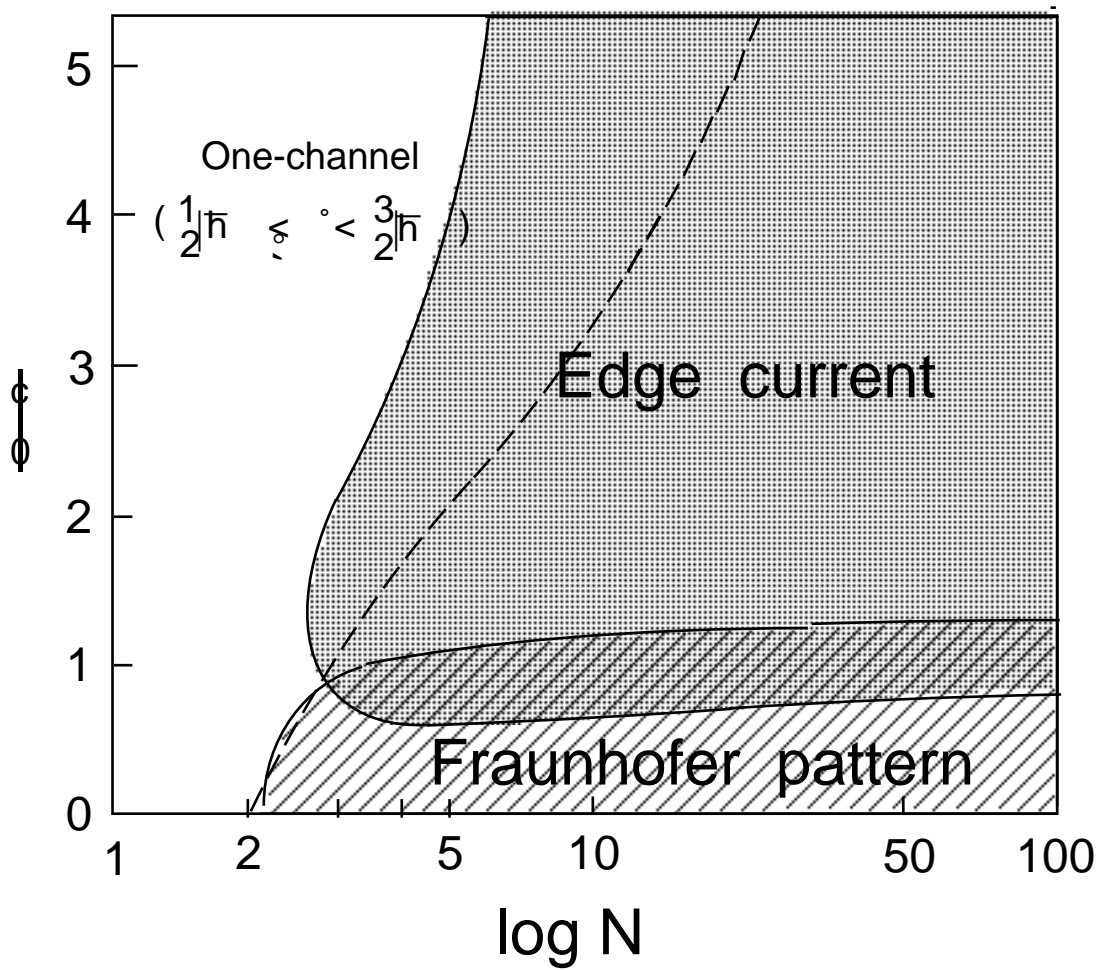
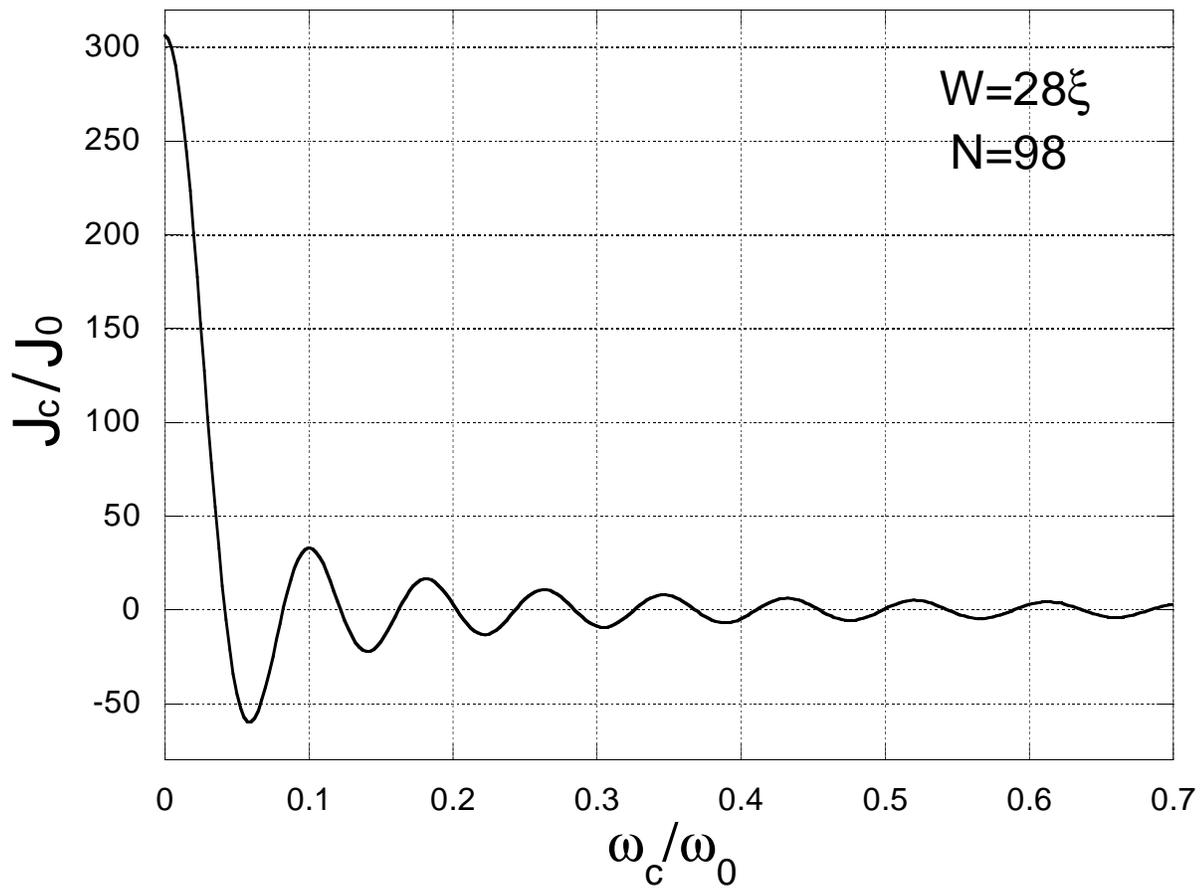
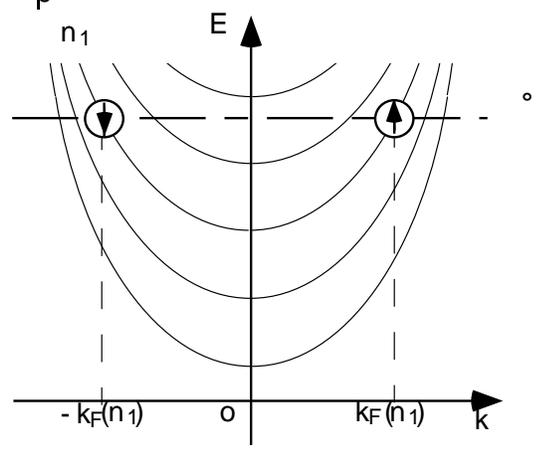
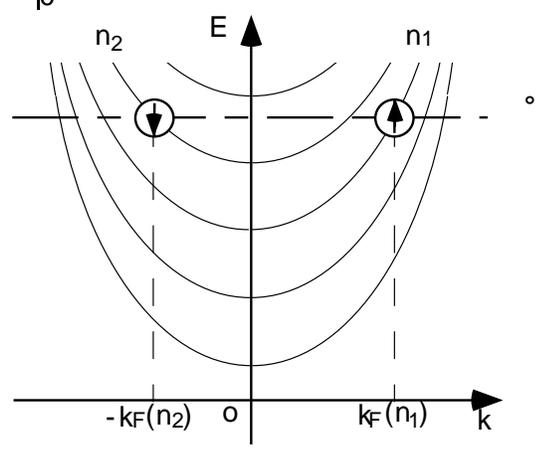
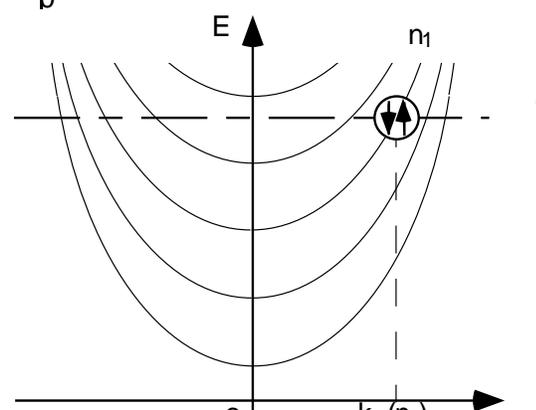
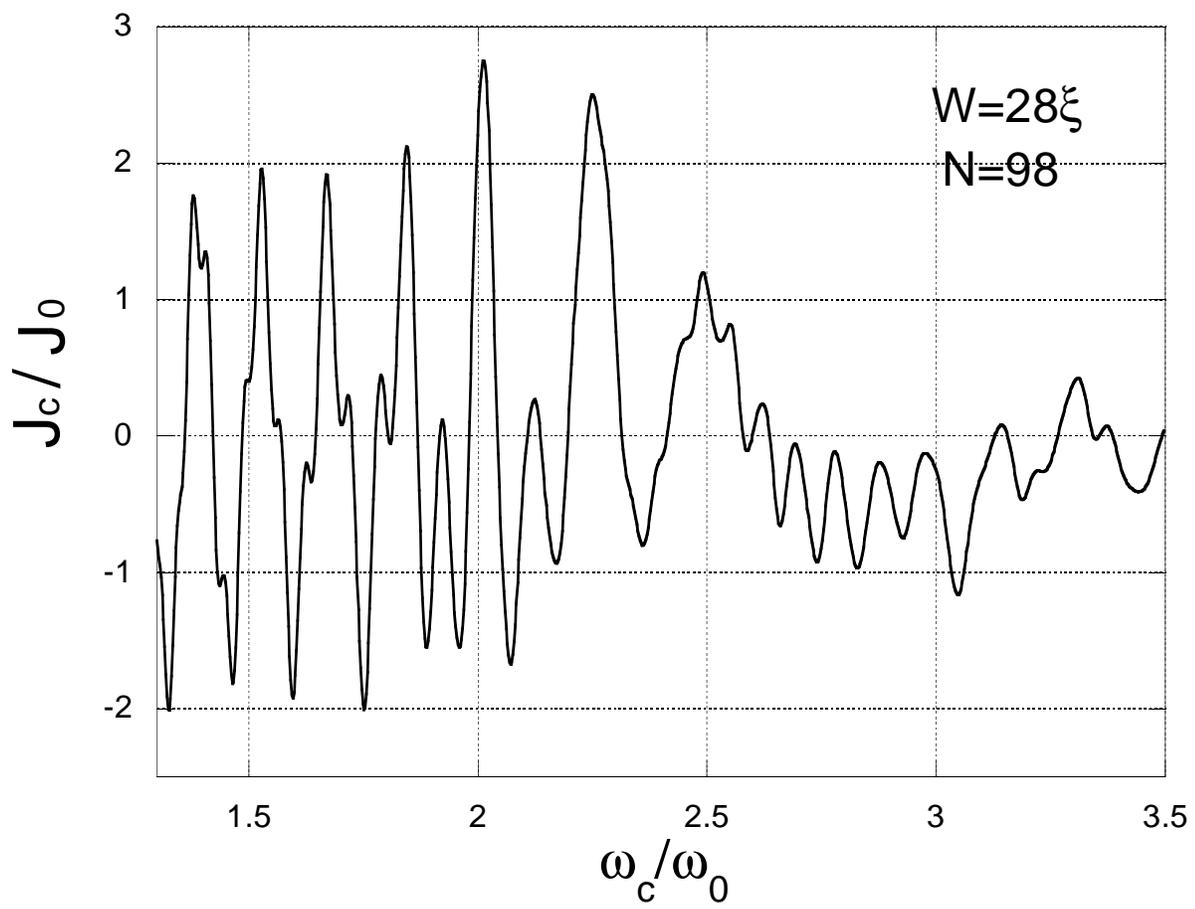
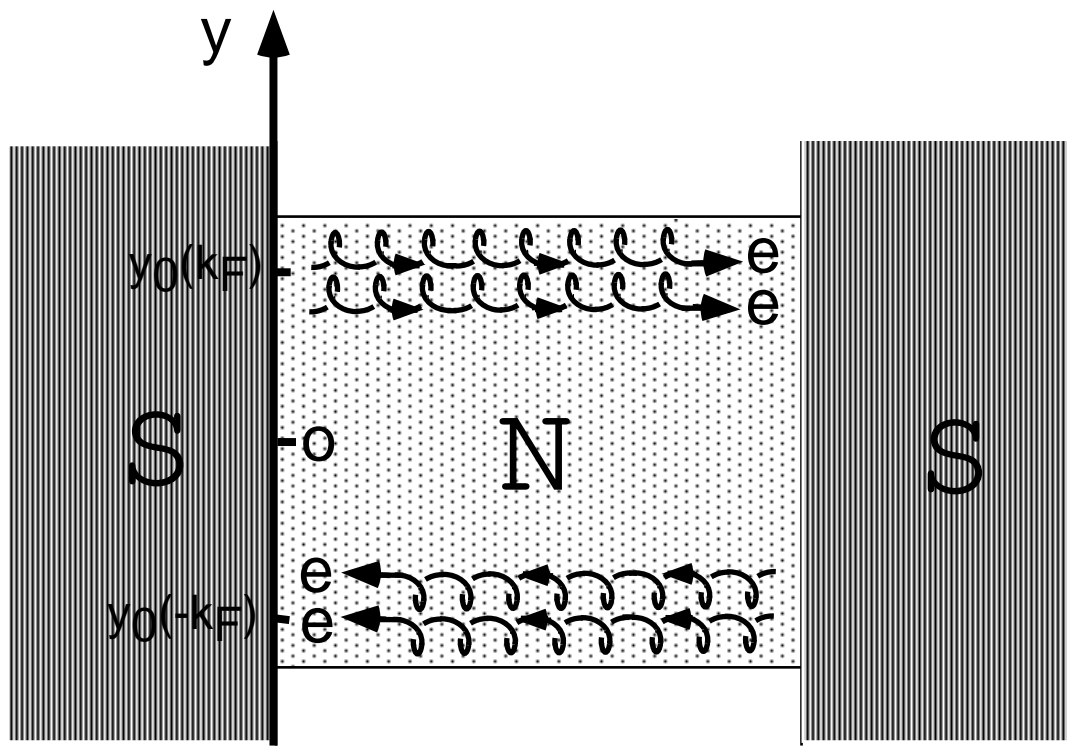
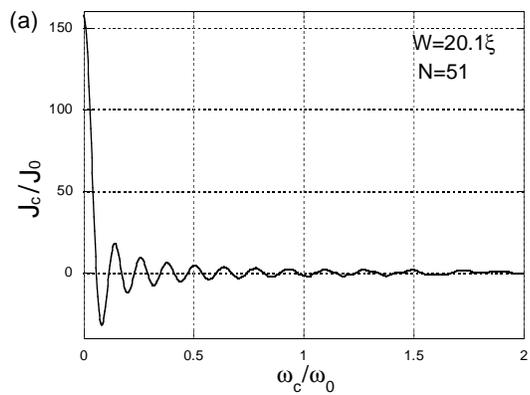
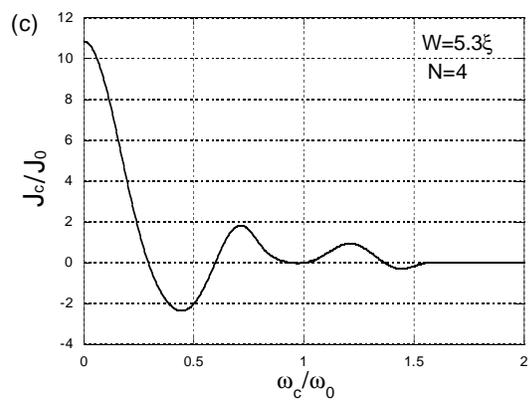
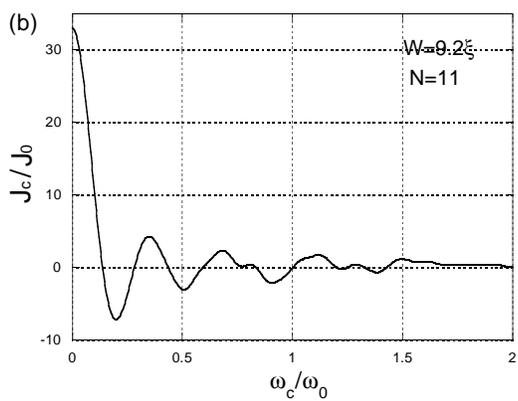
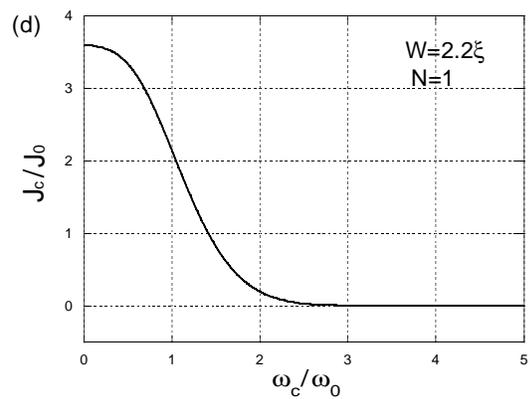
\end{document}